\documentclass[aps,prd,twocolumn,groupedaddress]{revtex4}
\usepackage{graphicx}
\newcommand{\be}{\begin{equation}}
\newcommand{\ee}{\end{equation}}
\newcommand{\ba}{\begin{eqnarray}}
\newcommand{\ea}{\end{eqnarray}}
\newcommand{\Eq}[1]{Eq.~(\ref{#1})}
\newcommand{\Eqs}[1]{Eqs.~(\ref{#1})}
\newcommand{\rf}[1]{(\ref{#1})}
\newcommand{\Ref}[1]{Ref.~\cite{#1}}

\newcommand{\Fig}[1]{Fig.~{\ref{#1}}}

\begin{document}
\title{Recent Advances in Dyson-Schwinger Studies\footnote{
Plenary talk given at NStar-2002, Pittsburgh PA, October 2002.}}
\author{M.A. Pichowsky}
\affiliation{Center for Nuclear Research and Department of Physics \\
Kent State University, Kent, OH 44242 U.S.A.}
\date{11 February 2003}
\begin{abstract}
  There have been many demonstrations of the utility of the
  Dyson-Schwinger equations of QCD as a systematic, phenomenological 
  framework for describing the perturbative and non-perturbative dynamics of
  hadrons in terms of Euclidean Green functions of quarks and gluons.
  Still, there remain some unanswered questions regarding the theoretical
  underpinnings of the approach.
  I review several studies that are shedding light on how such
  questions involving the connection between the DSEs and their use in
  phenomenological applications might be resolved, 
  and then review predictions for some exotic meson states.   
\end{abstract}
\maketitle
\section{Introduction}
%
The Dyson-Schwinger equations (DSEs) are an infinite set of coupled integral
equations that relate all of the dressed $n$-point functions of a quantum
field theory to each other.  
These $n$-point functions (which take their enumeration from the number of
space-time points they depend upon) determine all of the dynamics of
the theory.
With no external particle sources present, the most elementary of the
$n$-point functions of QCD are the dressed quark and gluon propagators
$S(x_1\!-x_2)$ and $D_{\mu\nu}(x_1\!-x_2)$.

A central aspect of QCD is that the elementary degrees
of freedom appearing in the Lagrangian (quarks and gluons) are {not}
simply related to the {observed} asymptotic degrees of 
freedom, the hadrons (mesons and baryons).  This is
referred to as {\em confinement} and understanding its origin represents
one of the outstanding theoretical problems in physics.
It then follows that knowledge of elementary two-point functions, the
dressed quark and gluon propagators might not, by itself, be very useful
since one can never observe quarks and gluons propagating directly.
Clearly, $n$-point functions such as the fully-interacting quark-antiquark 
$M(x_1,\ldots,x_4)$ and three-quark scattering amplitudes 
$H(x_1,\ldots,x_6)$, which are related to meson and baryon observables,
may be more interesting to explore.
However, the self-coupled nature of the Dyson-Schwinger equations entails
that these four-point and six-point functions are related to other
$n$-point functions including the two-point dressed quark and gluon
propagators $S(x_1\!- x_2)$ and $D_{\mu\nu}(x_1 \! - x_2)$.
Understanding the nature of this set of coupled, non-linear integral
equations is a daunting task, but advances are continually being made.
In the following, a summary of some of the most recent advances is
provided. 

One traces the study of the DSEs of QCD back to 1950s and 1960's.
These studies were crucial to understanding connections between field
theories and observables, but applications to hadron phenomena were only
moderately successful due to the lack of computational power available at
the time. 
In the 1980's, models formulated using the DSEs to describe the quark
dynamics within hadrons, were found to provide good and compact
descriptions of the light pseudoscalar and vector mesons. 
After these early phenomenological success, there followed a period of
theoretical studies of the DSEs which explored the role
of gluons, gauge invariance and confinement, which continues
to this day~\cite{Fischer,GluonDSE}.  

However, since the mid 1990's, much focus has turned to the 
study of physical processes which now permeates
almost every subfield of hadron physics including:    
chiral physics and pion decays~\cite{Review},
$\pi\pi$ scattering~\cite{Cotanch},
electromagnetic form factors of the light pseudoscalar~\cite{Tandy}
and light-vector meson octets~\cite{Hawes},
inelastic scattering and quark-parton distributions~\cite{Hecht}, 
high-energy diffractive electroproduction and hadron scattering~\cite{Lee},
finite temperature studies of QCD~\cite{Schmidt},
exotic mesons~\cite{BurdenPichowsky},
baryon bound states, their properties and electromagnetic form 
factors~\cite{Nucleon}, low-energy meson photo-production~\cite{Ahlig}, and
many others. 

Yet, a complete understanding of some theoretical aspects of the
framework has proven elusive.  
These unresolved questions include: 
How does one construct systematic
and convergent expansions for the kernels of the DSEs?
Are DSE and lattice QCD studies complementary approaches to the study of
QCD phenomena?
How does one calculate observables in Minkowski space from a quantum field
theory based in Euclidean space?
In the following, I give a brief account of recent studies that 
provide some answers to these questions.  
Finally, I report on an application of DSEs to exotic mesons.

\section{Robustness of Truncation Schemes}
 
The simplest DSE is the Dyson equation,   
\ba
\lefteqn{S^{-1}(p) = Z_2(\zeta,\Lambda) \, i \gamma\cdot p 
           + Z_{4}(\zeta,\Lambda)\, m(\zeta) }
\nonumber \\ && 
	   \!\!+ Z_{1}(\zeta,\Lambda)\!
           \int^{\Lambda}\!\!\!\!\! \frac{{\rm d}^4q}{(2\pi)^4}
             g^2 D_{\mu\nu}(p\!-q)
	   \frac{\lambda^{a}}{2}  \gamma_{\mu}   S(q)  
	   \Gamma^{a}_{\nu}(q,p),
	   \label{QuarkDse}
\ea
which gives the dressed quark propagator $S(p)$.
In \Eq{QuarkDse}, $\lambda^{a}$ for $a=1,\ldots,8$ are the 
Gell-Mann SU(3)-color matrices, $\Gamma^{a}_{\nu}(q,p)$ is
the dressed quark-gluon vertex and $m(\zeta)$ is the current quark mass.
A translation-invariant regularization scheme with scale $\Lambda$ is
employed to render the integral finite.
The dressed quark propagator is of the form, 
\be
   S(p) = \frac{Z(p^2)}{i \gamma \cdot p + M(p^2)},  
\ee
where $Z(p^2)$ is the quark wave-function renormalization and $M(p^2)$ is
the running quark mass.
The dependence of \Eq{QuarkDse} on the ultraviolet scale $\Lambda$ is
removed by renormalizing at $p^2=\zeta^2$, subject to the boundary conditions 
$Z(\zeta^2)=1$ and $M(\zeta^2)=m(\zeta)$,
and absorbing the $\Lambda$-dependence into the
multiplicative-renomalization constants $Z_{i}(\zeta,\Lambda)$. 

The dressed quark propagator is determined from 
\Eq{QuarkDse} only after the dressed gluon propagator $D_{\mu\nu}(p-q)$
and dressed quark-gluon vertex $\Gamma_{\nu}^{i}(q,p)$ are known.
However, these two- and three-point functions are solutions of their own
DSEs, which in turn depend on still {\em higher} $n$-point functions.
This is the infinitely-coupled nature of the DSEs, making it
impossible to obtain a {\em complete} solution of QCD directly.

In phenomenological applications, one may proceed by making assumptions
regarding some subset of $n$-point functions such that the DSEs are
reduced to a closed system of equations which may be solved directly. 
The simplest of such truncation schemes is the {\em rainbow approximation}.
It is equivalent to making the substitution, 
\ba
   Z_{1}(\zeta,\Lambda) \, g^2 \, \Gamma^{a}_{\nu}(q,p)
     \longrightarrow 
         \frac{\lambda^{a}}{2}  \gamma_{\nu} \, V\left((q-p)^2\right) , 
\label{Rainbow}
\ea
in \Eq{QuarkDse}, where $V(k^2)$ is an effective quark-gluon vertex.
The exact dressed quark-gluon vertex depends on eight distinct
Dirac-$\gamma$ matrices and on three variables $q^2$, $p^2$ and
$q\cdot p$, while the effective vertex in \Eq{Rainbow} maintains a single 
Dirac-$\gamma$ matrix and depends on $(q-p)^2$ only.

Assuming a given form for the dressed gluon propagator $D_{\mu\nu}(k)$ and
effective vertex $V(k^2)$, the quark propagator can be calculated 
from \Eq{QuarkDse}.
Then, one considers how a quark and antiquark combine to form a meson.  
This is described by the homogeneous Bethe-Salpeter equation (BSE), 
\be
\lambda(P^2) \Gamma(p;P) \!=\! \int^\Lambda \!\!\!\! 
       \frac{{\rm d}^4q}{(2\pi)^4}
       K(p,q;P)
       S(q_+) \Gamma(q;P)S(q_-)  , \;\;
\label{bse}
\ee
where $K(q,p;P)$ is the two-particle irreducible $q\bar{q}$ scattering
kernel,  $\Gamma(p;P)$ is the BS amplitude, and $q_\pm = q \pm P/2$
are the momenta of the quark and antiquark.
This can be solved for any value of meson momentum $P$, but when the
eigenvalue $\lambda(P^2)=1$, \Eq{bse} predicts the existence of a bound
state meson with mass $m=\sqrt{-P^2}$.   

One can show that when the rainbow approximation
is employed in \Eq{QuarkDse}, the ladder approximation for
the Bethe-Salpeter kernel, 
\be
  K(q,p;p) \rightarrow \frac{\lambda^{i}}{2} \gamma_{\mu} \otimes 
  \frac{\lambda^{i}}{2} \gamma_{\nu} \; D_{\mu\nu}(q-p) \, V((q-p)^2), 
  \label{Ladder}
\ee
must be employed in order to satisfy the axial-vector Ward-Takahashi
identity and Goldstone theorem~\cite{Bender}.
(A review may be found in \Ref{Review}.)
Furthermore, it can be argued that the rainbow-ladder approximation
represents the zeroth order term in a systematic series of approximations 
that preserve the Goldstone theorem.
However, what is truly amazing is that this series of truncations converges
at a surprisingly rapid rate~\cite{Bender,Detmold}!

To illustrate how such truncations might converge, one compares
solutions for the dressed quark propagator, and $\pi$- and $\rho$-meson
masses obtained using the ladder-rainbow truncations to those obtained
when higher-order terms are maintained in \Eqs{Rainbow} and \rf{Ladder}. 
In \Ref{Detmold}, this comparison was carried out using a simple model for
the dressed gluon propagator $D_{\mu\nu}(q-p)$.
The quark-gluon vertex was dressed by a one-gluon loop, two-gluon loops
and then with infinitely-many-gluon loops (using a recursive technique).  
The BSE kernel was constructed using the approach outlined in \Ref{Bender},
which ensured that the Goldstone theorem and axial-vector Ward-Takahashi identity
were maintained at each level of the truncation. 

The study demonstrates that $\pi$ and $\rho$ meson
bound state masses are well-described by the rainbow-ladder approximation
without significant change when dressing of the quark-gluon vertex is
included~\cite{Detmold}.
(In retrospect, this explains the success enjoyed by the many studies of
light pseudoscalar and vector mesons which have employed the ladder-rainbow
approximation.)
The study also found that deeper in the time-like region, the
dressing of the quark-gluon vertex may become more significant, 
but is still well-approximated by including only one-gluon or two-gluon loops;
that is, results for the quark propagator and the BSE 
obtained by using the recursively dressed quark-gluon vertex were very
similar to results obtained using just two-gluon loops to dress the
quark-gluon vertex~\cite{Detmold}.

The conclusion to be drawn from this study is that there is a systematic
method for improving truncation schemes employed in phenomenological
studies that allows a quantitative measure of their robustness. 
This achievement is quite important and represents a leap forward in
the use of Dyson-Schwinger equations as a phenomenological tool. 
However, there is no free ride.  Not all channels are equally robust or
insensitive to the truncation schemes.   
Rather, pseudoscalar and vector channels are particularly stable in this
respect. 
There are other channels for which the inclusion of dressings arising from
a higher number of gluons has significant impact.
Examples of such channels are the scalar and vector {\em colored-diquark}
channels~\cite{Bender,Detmold}.
Therefore, in practice one should check each channel of interest to verify
that results are stable under changes to the level of truncation employed.

\section{Connection to Calculation on the Lattice}
%
The similarities between studies of DSEs and numerical simulations of QCD
on the lattice are clear~\cite{Fischer,Maris,Jarecke}.
Both explore QCD from its Lagrangian in terms of quarks and gluon degrees
of freedom and both are formulated using the Euclidean metric where
the square of the momentum $p^2=p_{\mu} p_{\mu} = p\cdot p$ 
= $\sum_{i=1}^{4} p^2_{i} >0$ is space-like.
In principle, one might be able to compare results obtained from lattice
QCD simulations to those obtained from the DSEs directly.
In particular, lattice results may be used as input for DSE calculations or
might provide further tests of the robustness of truncation schemes.
Alternatively, Dyson-Schwinger studies may provide additional insights
into the role of some symmetries damaged by lattice QCD simulations;
for example, it is possible to use the Dyson equation~\rf{QuarkDse} 
to extrapolate lattice results for the quark propagator to the
chiral limit of vanishing quark mass.

\begin{figure}[t]
   \includegraphics[width=3.2in,viewport=0 0 700 525,clip=true]{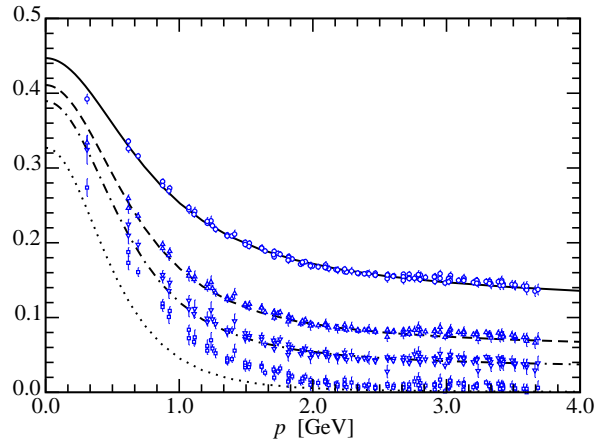}
   \caption{ 
   Comparisons of the running quark mass $M(p^2)$ obtained from 
   Eq.~\protect\rf{QuarkDse} for current quark masses $m(\zeta)=$ 
   27 (dot-dashed curve),  
   50 (dashed curve) and 102~MeV (solid curve) with corresponding results
   from lattice QCD simulations (data points)~\protect\cite{Bowman}.
   Also shown are $M(p^2)$ in the chiral limit obtained from
   the Dyson equation (dotted curve) and from a linear extrapolation of
   the lattice results (circles). 
   \label{Fig:LatticeM}
   }
 \end{figure}

This was carried out recently~\cite{BPRT} using a parametrization of the
Landau-gauge dressed gluon propagator $D_{\mu\nu}(k)$ obtained from a
lattice QCD simulation~\cite{Gluon}, and a simplified form of the dressed
quark-gluon vertex of the form \Eq{Rainbow} and parametrized by 
\be
V(k^2) = 
     \left({1 + \frac{a(m(\zeta))}{k^2}}\right)\left({1 +
       \frac{b}{k^2}}\right)^{-1}  
      + \cdots ,
\ee
where logarithmic perturbative-QCD corrections have been suppressed, 
$a$ and $b$ are parameters determined by using this vertex and the
lattice gluon propagator $D_{\mu\nu}(k)$ in \Eq{QuarkDse} to fit the
dressed quark propagator functions $Z(p^2)$ and $M(p^2)$. 
The resulting running quark mass functions $M(p^2)$ are compared to those
obtained from the lattice simulation~\cite{Bowman} for a collection
of current quark masses $m(\zeta)=$ 27, 50 and 102~MeV, 
as shown by the upper three curves in the left panel of \Fig{Fig:LatticeM}.
Self-consistency with lattice results required a dressed
quark-gluon vertex that depends on the current quark mass, 
$a(m)=\Lambda_{g}^2(1+7\,(m/\Lambda_g)$
$+125\, (m/\Lambda_g)^2)^{-1}$ and
$b=0.005~\Lambda_{g}^2$, where $\zeta=$ 19~GeV and $\Lambda_{g}=$
1020$\pm$100~MeV is a scale associated with the gluon propagator.

The agreement between the mass functions calculated on the lattice and from
the DSE \rf{QuarkDse} is excellent for all {\em finite} values of the current
quark mass $m(\zeta)$.
Having parametrized the dependence on the current quark mass, we can then
use this model to calculate the corresponding dressed quark propagator in
the chiral limit $Z_{4}(\zeta,\Lambda) \, m(\zeta) \rightarrow 0$.
The result is shown as the dotted curve in the left panel of
\Fig{Fig:LatticeM}. 
Because of numerical difficulties, it is impossible at present for the
lattice to simulate the chiral limit.  However, a simple, linear
extrapolation of the running quark mass function $M(p^2)$ may be
performed.  The result of such an extrapolation~\cite{Bowman} is depicted
by circles (lower data set) in \Fig{Fig:LatticeM}.
Clearly, the dotted curve fails to reproduce the linearly extrapolated
lattice results for $M(p^2)$.

This disagreement is caused by having employed too simple a form for the
dressed quark-gluon vertex in \Eq{Rainbow}, or from the inadequacy of a
{\em linear} extrapolation of lattice data to the chiral limit. 
Using the Gell-Mann--Oakes--Renner relation as a check on self-consistency
of the DSE approach suggests the latter may be true~\cite{BPRT}.
Reconciliation of these approaches requires additional investigation, but it
is clear that such comparisons are important and will provide new insights
for both Dyson-Schwinger and lattice communities in the future.

\section{Analytic Continuation into Minkowski Space}
%
Another issue, which has received less attention than those discussed above,
concerns the mechanism by which a quantum field theory, formulated
in Euclidean space, is related to hadron observables in Minkowski space.
Analytic continuation is the method by which observables in Minkowski space are
obtained from the Euclidean amplitudes or Green functions.
For example, the $\rho$-meson bound state mass $m_{\rho}$ corresponds to
finding the value of $P^2$ for which the BSE~(\ref{bse}) eigenvalue
$\lambda(P^2=-m_{\rho}^2)=1$.    
Since the Dyson-Schwinger framework is defined for Euclidean momenta
$P^2>0$, one must {\em analytically continue} the Bethe-Salpeter 
eigenvalue $\lambda(P^2)$ to time-like momenta $P^2=-m_{\rho}^2<0$. 

One way in which this process of analytic continuation may be understood
is by allowing the fourth component of the Euclidean vector $P_{\mu}$  to
become complex $P_4\rightarrow{iE}$ where $E>0$.
As a result of this continuation, during the integration over the (Euclidean)
relative-quark momentum $q_{\mu}$, the arguments of the quark propagators
$q_{\pm}^2=q^2+iE q_4-E^2/4$ sweep out a parabolic region of the complex
momentum plane, whose maximum width is proportional to the meson's energy
$E=m_{\rho}$.   
To accomplish this, one must know the behavior of the quark propagator
$S(q)$ within this parabolic region of the complex momentum plane. 
For light pseudoscalar ($\pi$ and $K$) mesons, the excursion into the
complex plane is minor and so is easy to handle.
For the light vector mesons ($\rho$, $\omega$, $K^*$ and $\phi$), 
this process can still be carried out by means of straight-forward
techniques, albeit numerically intensive~\cite{MarisTandy}.   
However, for mesons with masses significantly greater than 1~GeV, one may 
encounter poles and/or cuts in the quark propagator during this 
process of analytic continuation. 

It has long been thought that the dressed quark propagator may contain
complex-conjugate pairs of singularities and/or cuts in the
complex-momentum plane~\cite{Gribov}.
To explore how one might carry out an analytic continuation of the BSE
\rf{bse} in this event, we considered a simple model form for $S(q)$ given 
by $N$ pairs of complex-conjugate poles, 
\be
  S(q) = \sum_{n=1}^{N} \left(
        \frac{z_n}{i \gamma\cdot q+m_{n} } +
        \frac{z^{*}_{n}}{i \gamma\cdot q+m^*_{n}}
  \right),  
  \label{QuarkPoles}
\ee
where $m_n$ are complex-valued mass scales and $z_n$ are complex-valued
magnitudes~\cite{BPT}.

 \begin{figure}[t]
   \includegraphics[width=2.9in,angle=-90,clip=true]{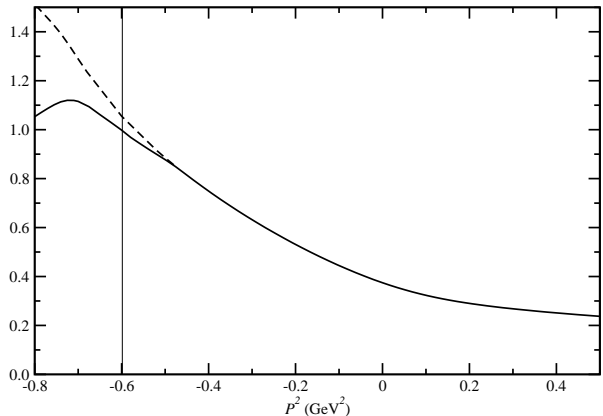}
   \vspace*{-2.0cm}
   \caption{ 
   The vector-meson ground state BSE eigenvalue $\lambda(P^2)$ for 
   a range of meson momenta $P^2$ (solid curve).
   The meson mass $m_{\rho}$ is calculated using 
   $\lambda(-m_{\rho}^2)=1$, as indicated by the vertical 
   line~\protect\cite{BPT}.
   \label{Fig:BSE}
   }
 \end{figure}

For the simple form given in \Eq{QuarkPoles} and a phenomenologically
successful BSE kernel~\cite{MarisTandy}, it was shown that one
can deform the contour integrations thereby avoiding quark singularities,
a necessary step to obtain the analytically continued eigenvalue
$\lambda(P^2)$ for {\em all} values of momenta 
$-\infty<P^2<+\infty$~\cite{BPT}.
The $P^2$ dependence of the resulting eigenvalue for the ground-state
vector meson is shown as a solid curve in the right panel of
\Fig{Fig:BSE}.  
For comparison, the dashed curve in \Fig{Fig:BSE} was obtained 
by completely ignoring the need to deform the contour around poles when
carrying out the integration in the BSE \rf{bse};  
as this is not the correct implementation of analytic continuation, the
resulting dashed curve is not correct.
The two curves are in perfect agreement for momenta 
$P^2>4{\rm Re}(m^2_1)$, where $m_1$ is the lightest quark mass parameter 
from \Eq{QuarkPoles}.   
For such momenta, no quark singularities are encountered during the
integration and so, the curves coincide here.
For momenta more time-like than this, the two curves diverge and the
solid curve is the correct analytically-continued eigenvalue
$\lambda(P^2)$.  
The ground state vector meson mass $m_{\rho}$ is the solution of 
$\lambda(P^2=-m_{\rho}^2)=1$. 

The Dyson-Schwinger framework is a renormalizable, Euclidean
approach to the study of QCD.  As such, one makes contact with observables
by analytically continuing obtained amplitudes into Minkowski space.
Using recently developed numerical methods, one can perform analytic
continuations of the BSE, even in the presence of singularities in the
quark propagators~\cite{BPT}.  
We now have the means to explore meson and baryon bound states of much
heavier masses than previously possible.

\section{Exotic Mesons from the Bethe-Salpeter Equation}
%
Before summarizing, I provide one example of a simple model calculation of
the meson spectrum.  
Studies of hadrons (both baryons and mesons) within the Dyson-Schwinger
framework are too numerous to describe adequately in the space provided
here. 
An extensive review, appearing recently, provides an excellent summary of
the current status of such studies~\cite{Review}.
Rather than reiterate aspects of these studies, I give a brief accounting
of a recent study of {\em exotic mesons} not covered therein.  

 \begin{figure*}[t]
\begin{center}
  \includegraphics[width=2.5in,clip=true,angle=-90]{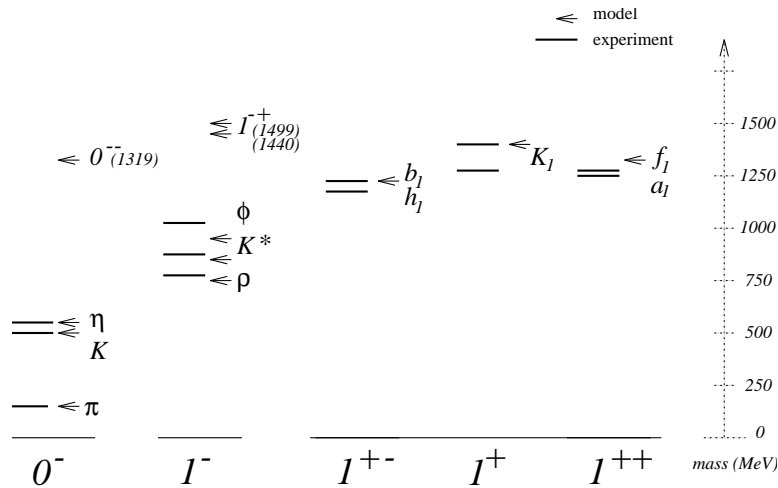}
  \end{center}
\vspace*{-0.30cm}
\caption{Meson spectra obtained in  Refs.~\protect\cite{BurdenPichowsky} 
  and \protect\cite{BQRTT}.
  One exotic pseudo-scalar with mass of 1319~MeV and two exotic vector
  mesons with masses of 1440 and 1499~MeV are obtained from the model.
 \label{Fig:BQRTT}
}
 \end{figure*}

The most famous of exotic mesons is probably the $\pi_1(1400)$, 
the iso-vector, vector meson with positive charge-conjugation parity and
quantum numbers $J^{PC}=1^{-+}$.
The classification of ``exotic'' belongs to those mesons with quantum
numbers that can not be constructed as a quark-antiquark \underline{state}.
A quark-antiquark state $|q\bar{q}\rangle$ with relative spin $S$ and
angular momentum $L$ will have a spatial parity $P=(-1)^{L+1}$ and
charge-conjugation parity of $C=(-1)^{L+S}$.  
To generate such a state, one must include additional degrees of freedom.  
In the quark model, this may be accomplished by introducing either
constituent gluons or flux-tube excitations or perhaps, by considering 
composite systems like $|\bar{q}\bar{q}qq\rangle$.

In a quantum field theoretic framework, such particles are generated by
constructing operators with the desired $J^{PC}=1^{-+}$ 
quantum numbers.
One such operator is $\bar{\psi}(x)\vec{\gamma}D_{4}\psi(x)$ where
$\psi(x)$ is the quark field and $D_{4}$ is the fourth component of the 
covariant derivative
$D_{\mu}=\partial_{\mu}+igA_{\mu}$~\cite{BurdenPichowsky}. 
The first part of this operator may be used to project out the exotic meson
bound state solution from the Bethe-Salpeter equation.  Predictions for
the masses of two exotic vector mesons were obtained~\cite{BurdenPichowsky}
using the simple separable model~\cite{BQRTT} which provides an excellent
description  of the light pseudoscalar, vector and axial-vector meson
spectra as shown in \Fig{Fig:BQRTT}.
One exotic pseudoscalar and two exotic vector mesons are predicted.

\section{Summary}
Recent theoretical advances are putting the Dyson-Schwinger framework on 
firmer ground and in closer contact with complementary frameworks,
such as lattice QCD simulations. 
Future studies will continue to increase our ability to use this
framework as a phenomenological tool for exploring the
dynamics of quarks and gluons in hadron processes.

{\bf Acknowledgements: }
This work is supported by the National Science Foundation under contract
number \uppercase{PHY}-0071361. 
Some original work reported herein resulted from collaborations with 
M.S.~Bhagwat and P.C.~Tandy.


\end{document}